\documentclass[submission,copyright]{eptcs}
\providecommand{\event}{DCM 2012} 
\usepackage{breakurl}             

\usepackage{graphicx}
\usepackage{amssymb,amsmath,amsthm}
\allowdisplaybreaks
\def\nyil{\rightarrow}
\def\Values{\mathbb{V}}
\def\Reals{\mathbb{R}}

\newtheorem{definition}{Definition}
\newtheorem{lemma}[definition]{Lemma}
\newtheorem{theorem}[definition]{Theorem}

\def\beginDefinition{\begin{definition}}
\def\endDefinition{\end{definition}}
\def\beginTheorem{\begin{theorem}}
\def\endTheorem{\end{theorem}}
\def\beginLemma{\begin{lemma}}
\def\endLemma{\end{lemma}}
\def\beginProof{\begin{proof}}
\def\endProof{\end{proof}}

\renewcommand{\emptyset}{\varnothing}

\newcommand{\AND}{\mathrm{AND}}
\newcommand{\OR}{\mathrm{OR}}
\newcommand{\PRODUCT}{\mathrm{PRODUCT}}
\newcommand{\RSHIFT}{\mathrm{RSHIFT}}
\newcommand{\LSHIFT}{\mathrm{LSHIFT}}
\newcommand{\OUTPUT}{\mathrm{OUTPUT}}
\newcommand{\NOT}{\mathrm{NOT}}
\newcommand{\FIRSTHALF}{\mathrm{FIRSTHALF}}
\newcommand{\RIGHT}{\mathrm{RIGHT}}
\newcommand{\Lshift}{\mathrm{Lshift}}
\newcommand{\rmfrac}{\mathrm{frac}}
\newcommand{\Rshift}{\mathrm{Rshift}}
\newcommand{\Flength}{\mathrm{Flength}}

\newcommand{\rmod}{\mathrel{\mathrm{mod}}}

\title{Computing discrete logarithm by interval-valued paradigm}
\author{Benedek Nagy
\institute{Faculty of Informatics\\University of Debrecen, Hungary}
\institute{Department of Mathematics\\Eastern Mediterranean University, Turkey}
\email{nbenedek@inf.unideb.hu}
\and
 S\'{a}ndor V\'{a}lyi
\institute{Institute of Mathematics and Informatics\\College of Ny\'{\i}regyh\'{a}za,
Hungary}
\email{valyis@nyf.hu}
}
\def\titlerunning{Computing discrete logarithm by interval-valued paradigm}
\def\authorrunning{B. Nagy \& S. V\'{a}lyi}
\begin{document}
\maketitle

\begin{abstract}
Interval-valued computing is a relatively new computing paradigm. It uses finitely many interval
segments over the unit interval in a computation as data structure.
The satisfiability of Quantified Boolean formulae and other hard problems, like integer factorization,
can be solved in an effective way by its massive parallelism.
 The discrete logarithm problem plays an important
role in practice, there are cryptographical
methods based on its computational hardness.
In this paper we show that the discrete logarithm problem
is computable by an interval-valued computing in a polynomial number of steps (within this paradigm).
\end{abstract}

\section{Introduction}

There are intractable problems that  traditional computing devices (Turing machines, Neumann architecture computers) cannot solve
efficiently. For some classes of hard problems, such as NP-complete, PSPACE etc., it is strongly believed that one cannot find
any method to solve these in deterministic polynomial time. Such a hard problem is to find the discrete logarithm of a
given positive integer. Several cryptographic methods are based on the assumption that the computation of discrete logarithm
cannot be achieved in deterministic polynomial time  \cite{crypto}.

There are various new theoretical computing paradigms \cite{parad} that attack these hard problems successfully at least in theory.
There are various paradigms based on inspiration from Biology (e.g., DNA-computing, membrane computing), from Physics (e.g., Quantum computing)
and from other phenomena of the Nature.
The efficiency of most of these new paradigms come from a massive parallelism built in the system (the power of quantum computation also derives from entanglement). In this paper we dealt with another new paradigm that uses strong inner parallelism.
%
%
The Interval-valued computing paradigm uses finitely many interval
segments over the unit interval in a computation.
Logical operations are straightforward generalizations of the classical bit
operations (of usual computers), moreover shift operations is also used to carry some pieces
of information to other parts of the unit interval. The product operation
allows to raise the density of information.
The paradigm has
been investigated in \cite{CiE2005,VLL2007} as a way of visual
computations and proved to be very efficient, e.g., PSPACE is
characterized by restricted polynomial interval-valued
computations in \cite{TCS2008}. Computationally hard problems are
solved in efficient way in this paradigm, e.g., the PSPACE complete problem of satisfiability of Quantified Boolean formulae in
\cite{TCS2008}, integer prime factorization in \cite{publi2011}.

In this paper we solve another computationally hard problem,
namely the discrete logarithm problem, that plays an important
role both in mathematical theory and in practice, e.g., it is related to some cryptographical
algorithms \cite{dl,crypto}. We note here that other new computing paradigms also
address this problem, see, e.g., \cite{Shor}, where a Quantum algorithm is presented
to solve the discrete logarithm problem efficiently.


The structure of the paper is as follows.
In the next section we recall the interval-valued paradigm in a formal
way. In Section 3 our algorithm is presented that solves the discrete logarithm problem
in a cubic complexity,
finally in Section 4  some concluding remarks close the paper.

\section{Preliminaries}
For the sake of a self-contained paper, we repeat the needed definitions from \cite{TCS2008} and \cite{publi2011}.
 First we define what  an interval-value means. Then we
present the allowed operations which can be used to build and
evaluate computation sequences.
We also give the notions
concerning decidability and computational complexity.

\subsection{Interval-values}

We note in advance that we do not distinguish interval-values (specific functions from [0,1) into $\{0,1\}$) from their subset representations \ (subsets of [0,1)) and  we always use the more convenient notation.

\beginDefinition\label{interval-values} The set $\Values$ of
\emph{interval-values} coincides with
the set of finite unions of $[)$-type subintervals of $[0,1)$.
\endDefinition

\beginDefinition\label{specific-interval-values} The set $\Values_0$ of
\emph{specific interval-values} coincides with
\begin{equation}\label{eq:star}
\left\{\bigcup\limits_{i=1}^{k}
\left[\frac{l_i}{2^m},\frac{l_i+1}{2^m}\right)\ : m \in \mathbb{N},
k \leq 2^m, 0 \leq l_1 < \ldots < l_k < 2^m \right\}.
\end{equation}
Similarly, let $\Values_n$ be the set of interval-values that can be represented
by (\ref{eq:star}) using only values $m$ with
the condition
$m\leq n$. %
\endDefinition

We note that the set of finite unions includes the empty set
$(k=0)$, that is, $\emptyset$ is also an allowed interval-value.

\subsection{Operators on interval-values}

 If we consider interval-values
as subsets of [0,1), then the set-theoretical operations such as complementation 
($\overline{A}$), union ($A \cup B$) and intersection ($A \cap
B$) on $\Values$ are definitely applicable. The algebra $(\Values,{}\bar{} {},\cup,\cap)$ forms an infinite Boolean set algebra with these
operations, $\Values_0$ is one of its infinite subalgebra, while the systems
based on $\Values_n$ $(n>0)$ are finite subalgebras.

\beginDefinition\label{firstlength}
The first component of an interval value $A \in \Values$, $A\ne \emptyset$, is defined as the interval value $[t,s)$
where $t$ and $s \in [0,1]$ satisfy that $[0,t) \cap A = \emptyset$, $[t,s) \subset A$ and $\forall s'>s: [t,s') \not\subset A$.
Now the function
$\Flength: \Values \rightarrow \Reals$ is defined as follows.
If $A = \emptyset$, then $\Flength(A) = 0$. Otherwise
 $\Flength(A)= s-t$, where $[t,s)$ is the first component of $A$.
   \endDefinition

Intuitively, this function provides the length of the left-most component
(included maximal subinterval) of an interval-value $A$.
The function $\Flength$ helps us to define the binary shift operators on
$\Values$. The \emph{left-shift} operator will shift the first
interval-value to the left by the first-length of the second
operand and remove the part which is shifted out of the interval
$[0,1)$ to the negative direction. As opposed to this,  the \emph{right-shift} operator is
defined in a circular way, i.e., the parts shifted above 1 will
appear at the lower end of $[0,1)$. In this definition we write
interval-values in their  ``characteristic function''
notation.

\beginDefinition\label{shifts} The binary operators $\Lshift$ and $\Rshift$ on $\Values$ are defined in the following way.
If $x \in [0,1)$ and $A, B \in \Values$ then
$$\Lshift(A,B)(x) = \left\{
\begin{array}{ll}
A(x + \Flength(B)) & \ \textrm{if}\ 0  \le x + \Flength(B) < 1,\\
0\ & \ \textrm{in other cases;} \end{array}\right. $$
and
$$\Rshift(A,B)(x) = A(\rmfrac(x - \Flength(B))),$$
where the function $\rmfrac$ gives the fractional part of a real number, i.e., $\rmfrac(x)=x - \lfloor x\rfloor$, where $\lfloor x\rfloor$ is the greatest integer which is not greater than $x$. \endDefinition

By the combined application of the shift operators we can choose any `important' part
of the interval-values by erasing its complement.

\beginDefinition
Let $A$ and $B$ be
 interval-values and $x \in [0,1)$. Then the  (fractalian) product $B * A$ includes $x$ if and only if $B(x) = 1$ and $A\left(\frac{x-x_B}{x^B-x_B}\right)=1$, where $x_B$ denotes the lower end-point of the $B$-component including $x$, and $x^B$ denotes the upper end-point of this component, that is, $[x_B, x^B)$ is the maximal subinterval of $B$ containing $x$.
\endDefinition

We can give this operation in a more descriptive manner. If $A$ contains $l$ interval components with end points $a_{i,1} , a_{i,2}$ $
(1 \leq i \leq l)$ and $B$ contains $k$ components with end points
$b_{i,1}, b_{i,2} $ $(1 \leq i \leq k)$, then we determine the
value of $C = B*A$ as follows: we set the number of components of $C$ to be $l \cdot k$. For this process we can use double indices for the components of $C$.
The lower and higher %
end-points of the
$(i,j)$th component are $b_{i1}+a_{j1}(b_{i2}-b_{i1})$ and
$b_{i1}+a_{j2}(b_{i2}-b_{i1}), $ respectively. In visual way, the interval-value $A$ is
zoomed to the components of the interval-value $B$.
   The idea and the role of this operation is similar to the
unlimited shrinking of 2-dimensional images in optical computing. %
 It will be used
to connect interval-values of different resolution (i.e, increase $n$ in the actually used
$\Values_n $).

\bigskip

\subsection{Syntax and semantics of computation sequences}
 This formalism is of Boolean network style. As usual,
the length of a sequence $S$ is denoted by $|S|$ and its $i$th
element by $S_i$. If $j \leq |S|$ then the $j$-length prefix of
$S$ is denoted by $S_{\nyil j}$.

\beginDefinition\label{syntax} An
\emph{interval-valued computation sequence} is a nonempty finite
sequence $S$ satisfying $S_1 = \FIRSTHALF$ and further, for any $i
\in  \{2,\ldots,|S|\}$, $S_i$ is $(\mathrm{op},l,m)$ for some $$\mathrm{op} \in
\{\AND,\OR,\LSHIFT,\RSHIFT,\PRODUCT\}$$ or $S_i$ is
$(\NOT,l)$ or $(\OUTPUT,l)$ where $\{l,m\}\subseteq\{1,\ldots,i-1\}$.

One of the complexity measures of a given computation is the \emph{bit height} of a computation. It is the minimal value $n$ such that all the interval-values of the computation are
in $\Values_n$.
\endDefinition

The semantics of  interval-valued computation  sequences is defined by induction on the length of the sequences. The \emph{interval-value} of such a sequence $S$ is denoted by $\|S\|$ and defined by induction on the length of the computation sequence, as follows.

\beginDefinition\label{semantics}  First, we fix
$\|(\FIRSTHALF)\|$ as $\left[0,\frac{1}{2}\right)$. Second, if $S$
is an interval-valued computation sequence and $|S|$ is its length, then
$$\|S\|=\left\{ \begin{array}{ll}
\|S_{\nyil j}\| \cap \|S_{\nyil k}\|,\ & \textrm{if }\ S_{|S|}=(\AND,j,k), \\
\|S_{\nyil j}\| \cup \|S_{\nyil k}\|,\ & \textrm{if }\ S_{|S|}=(\OR,j,k)\\
\|S_{\nyil j}\| * \|S_{\nyil k}\|,\ & \textrm{if }\ S_{|S|}=(\PRODUCT,j,k)\\
\Rshift(\|S_{\nyil j}\|,\|S_{\nyil k}\|),\ & \textrm{if }\ S_{|S|}=(\RSHIFT,j,k)\\
\Lshift(\|S_{\nyil j}\|,\|S_{\nyil k}\|),\ & \textrm{if }\ S_{|S|}=(\LSHIFT,j,k)\\
\|S_{\nyil j}\|,\ & \textrm{if }\  S_{|S|}=(\OUTPUT,j)\\
\overline{\|S_{\nyil j}\|},\ & \textrm{if }\ S_{|S|}=(\NOT,j).
\end{array} \right.$$
\endDefinition

Here the system of \cite{TCS2008} is
 extended with an instruction to write (i.e., generate) the output as we detail below
 based on 
  \cite{publi2011}.

\subsection{Computing a discrete function by interval-values}
The semantics of writing the output is the following.
The output sequence is an element of $\{0,1\}^*$, initially the empty sequence.
Let $S_1\ldots S_n$ denote the computation sequence.
If $S_j = (OUTPUT,i)$ where $i<j$ then $\|S_{\nyil j}\| = \|S_{\nyil i}\|$ and
 as a side effect, 1 is concatenated to  the output sequence if $S_j$ is nonempty,
 otherwise 0 is concatenated to it. The answer of a computation sequence is its output sequence produced
during the computation.
Let $f:\{0,1\}^*\rightarrow \{0,1\}^*$. We say that $f$ is computable by an interval-valued computation
if and only if there exists a logspace algorithm $\mathcal B$
that for each possible input ($w \in \{0,1\}^*$) constructs
a computation sequence that generates the  output sequence $f(w)$.

The \textit{size of a computation} is measured by the length of the computation sequence.
We recall from \cite{TCS2008} that the class of polynomial size interval-valued computations
in which one of the arguments of every product operation is $\FIRSTHALF$
characterizes the classical complexity class PSPACE.

\section{Computing the discrete logarithm by interval-values}

In this section we solve the discrete logarithm problem within the interval-valued paradigm. Let the input $a,b,p\in \mathbb Z$. 
We give an
interval-valued computation sequence that give the result as output: an exponent $x$
of input integer  $a$ such that %
  $a^x = b \rmod p$ holds. We can assume without loss of generality that $a$, $b$ and
$x$ are non-negative integers less than $p$.

\beginTheorem\label{fotetel}{Discrete logarithm can be computed  by an interval-valued computation of size $O(n^3)$.}
\endTheorem

We prove this theorem in a constructive
way through several Lemmas in this section.

 The computation of discrete logarithm usually means the following computing task: For any input triplet$(a,b,p)$, where $p$ is a prime,
$a$ and  $b$ are non-negative integers less than $p$, find a non-negative integer $x$
such that $a^{x}=b \rmod p$ holds. There is a value $x$ such that $x<p$, therefore our
search will check only integers that can be represented at most as many bits as $p$ can be.
Throughout in this paper we denote  the upper integer part of the usual logarithm of $p$ by $n$.
That is, $a$, $b$, $x$ and $p$ all can be written by $n$ binary digits.
 Let $a_1\ldots a_n$ be the binary representation of the input
integer $a$. (One can assume that $n\geq 3$.) Similarly, $b_1\ldots b_n$ is the binary form of $b$ and
$p_1\ldots p_n$ is of $p$.
 We give a logspace algorithm $\mathcal B$ that constructs an
interval-valued computation sequence $S$ from $a, b$ and $p$ with an output
bit sequence $d_1\ldots d_n$ that is the binary representation of
the target $x$.

The algorithm $\mathcal B$ starts its work by representing the input bit sequences ($a$, $b$ and $p$) by interval-values. First, fix $S_1$
as $\FIRSTHALF$ and $S_2$ as $(\RIGHT,1,1)$. Then, for each $i \in \{1,\ldots,n\}$, if $a_i = 1$, then put $S_{2+i}:=(\OR,1,2)$ else  $S_{2+i}:=(\AND,1,2)$;
if $b_i = 1$, then put $S_{2+n+i}:=(\OR,1,2)$ else  $S_{2+n+i}:=(\AND,1,2)$; and
if $p_i = 1$, then put $S_{2+2n+i}:=(\OR,1,2)$ else  $S_{2+2n+i}:=(\AND,1,2)$.
We denote the indices of the subsequence $S_3,S_4,\ldots,S_{2+n}$
by $a(1),a(2),\ldots,a(n)$. Indices $b(1),b(2),\ldots,b(n)$ and $p(1),p(2),\ldots,p(n)$ can be defined similarly.
In this way we have represented the input:

\beginLemma For each $k \in \{1,\ldots,n\}$:
\begin{align*}
\|S_{\nyil a(k)}\| &=
\left\{ \begin{array}{ll}[0,1) & \ \textrm{if}\ a_k = 1 \mbox{ and}\\
\emptyset &\ \textrm{if}\ a_k = 0;\end{array}\right.\\
\|S_{\nyil b(k)}\| &=
\left\{ \begin{array}{ll}[0,1) & \ \textrm{if}\ b_k = 1 \mbox{ and}\\
\emptyset &\ \textrm{if}\ b_k = 0;\end{array}\right. \\
\|S_{\nyil p(k)}\| &=
\left\{ \begin{array}{ll}[0,1) & \ \textrm{if}\ p_k = 1 \mbox{ and}\\
\emptyset &\ \textrm{if}\ p_k = 0.\end{array}\right.
\end{align*}
\endLemma

\beginProof
This is straightforward. \endProof

\begin{figure}
\centering
\includegraphics[width=0.45\textwidth]{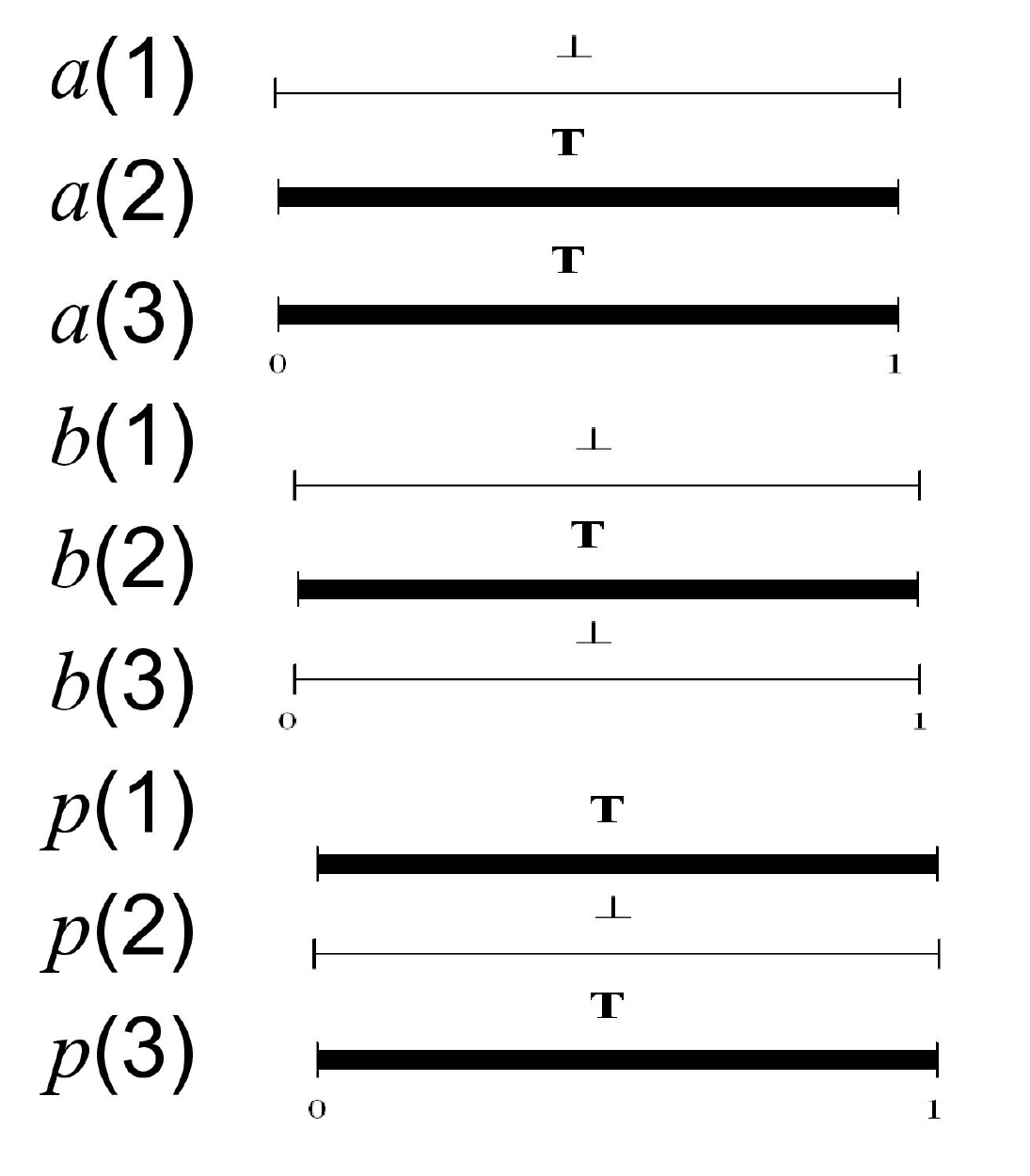}
\caption{Representation of the input by interval-values.}\label{fig:input}
\end{figure}

We illustrate our algorithm by an example: $a=3$, $b=2$ and $p=5$. In Figure \ref{fig:input} an illustration is given for the initialization part of the algorithm.

All possible candidates for $x$ are going to be represented in a parallel way in different slices of the interval values.
$\mathcal B$ continues its job 
by computing $S_{p(n)+1},\ldots,S_{p(n)+(3n-2)}$ as follows: $S_{p(n)+1} = (\AND,1,1)$.
For all positive integers $k < n$,
\begin{align*}
S_{p(n)+3k-1} & =(\PRODUCT,p(n)+3k-2,1),\\
S_{p(n)+3k}&=(\RSHIFT,p(n)+3k-2,p(n)+3k-1)\mbox{ and}\\
S_{p(n)+3k+1}&=(\OR,p(n)+3k,p(n)+3k-1).
\end{align*}

The index sequence $p(n)+1,p(n)+4,\ldots,p(n)+(3n-2)$ will be denoted by $x(1),x(2),\ldots,x(n)$.
By induction on $k$ one can establish the following statement.

\beginLemma
For all integer $k \in \{1,\ldots,n\}$:
$$\|S_{\nyil x(k)}\| = \|S_{\nyil p(n)+(3k-2)}\| = \bigcup\limits_{l=0}^{2^{k-1}-1}
\left[\frac{2l}{2^{k}},\frac{2l+1}{2^{k}}\right).$$
\endLemma

Similar representations were used, for instance, in \cite{TCS2008} for a concise representation of all the possible evaluations of a Boolean formula.
In this way all variations of
$n$ independent bits can be represented simultaneously by the interval-values
$\|S_{\nyil x(1)}\|,$ $\|S_{\nyil x(2)}\|,$ $\ldots,\|S_{\nyil x(n)}\|$ in the following sense:

\beginLemma\label{mindenvariacioszerepel}
For each bit sequence $t_1\ldots t_{n}$  there exists $r \in [0,1)$ that
for any $k \in \{1,\ldots,n\}$: $r \in \|S_{\nyil x(k)}\|$ if and only if $t_k = 1$.
\endLemma

\begin{figure}
\centering
\includegraphics[width=0.35\textwidth]{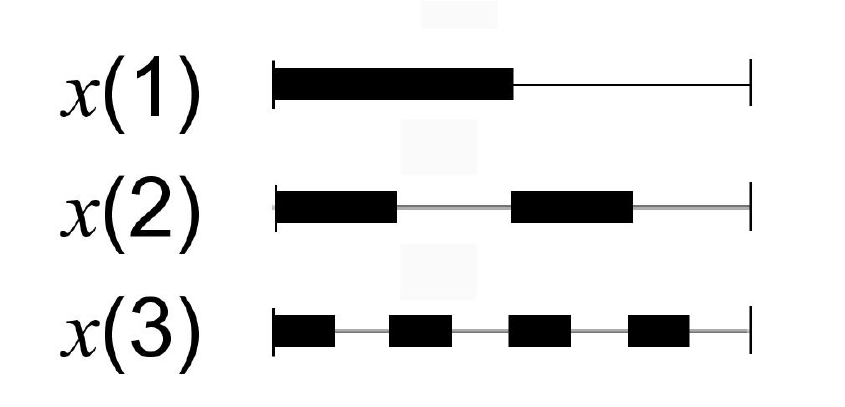}
\caption{Representation of the possible solutions.}\label{fig:x}
\end{figure}

The choice $r = \sum\limits_{i=1}^{2n}
\frac{1-t_i}{2^i}$  proves the lemma. 
In our example the largest number is 5 and it can be represented on 3 bits, therefore the case $n=3$ is visualized in Figure \ref{fig:x}.
Now a further definition is needed to find the coded (represented) values.

\beginDefinition
For $k \in \{1,\ldots,n\}$ and $r \in [0,1)$, let  $x_k(r):= (r \in \|S_{\nyil x(k)}\|)$. Further, let the bit sequence
$x_1(r)\ldots x_n(r)$ be denoted by $X(r)$. For any bit sequence
$BS = b_1\ldots b_n$, let $\#BS$ denote the integer whose binary
representation is $BS$.
\endDefinition

Let us construct a Boolean circuit of size $m+n$ ($m>0$) that multiplies two $n$-length input
bit sequences (interpreted as integers in binary form) modulo a third
$n$-length input bit sequence outputting the $i$th output bit in step $m+i$ ($i \in \{1,\ldots,n\}$).
The circuit can be chosen so that circuit size $m$ will depend on $n$ quadratically.

It can be simulated by an interval-valued computation sequence using only the corresponding Boolean operators.
Let $e(i,j)$ abbreviate $x(n)+(i-1)\cdot n+i \cdot m+j$, for any $(i,j) \in \{1,\ldots,n\}^2$.
Applying the chosen multiplier computation sequence $n$ times to the appropriate operands, $\mathcal{B}$ can construct an interval-valued
computation sequence that satisfies the formula
\[
\|S_{\nyil e(i,j)}\| =
\left\{
\begin{array}{ll}
[0,1) , & \textrm{ if the} j\textrm{th bit of } (a^{2^i} \rmod p) \textrm{ is } 1, \\
\emptyset , & \textrm{ otherwise};
\end{array}
\right. \]
for any $(i,j) \in \{1,\ldots,n\}^2$.

\begin{figure}
\centering
\includegraphics[width=0.9\textwidth]{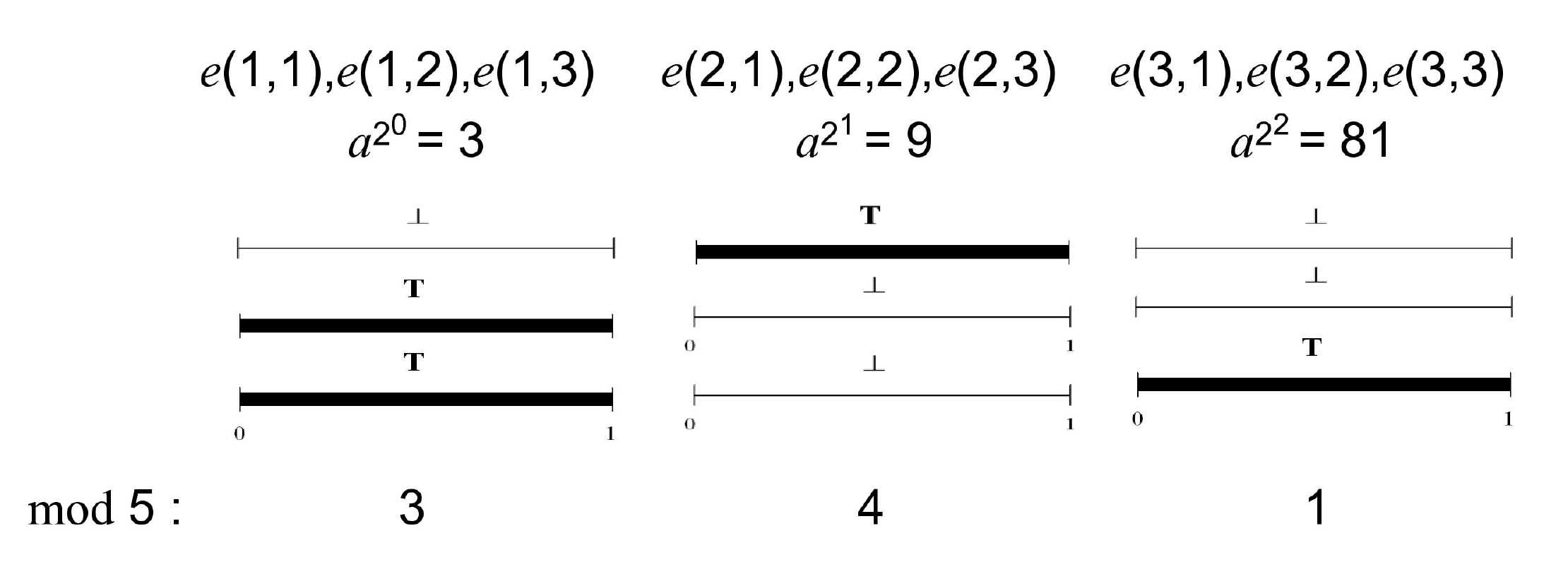}
\caption{The values $e(i,j)$ with $a=3$, $b=2$ and $p=5$.}\label{fig:e}
\end{figure}

The length of the actual part of the constructed computation sequence is in $O(n^3)$.
Figure \ref{fig:e} shows the interval-values $\{e(i,j) | 1 \leq i,j \leq 3\}$ in our example.

The algorithm $\mathcal{B}$ continues to build the computation sequence. This part is of length $3n^2$ and ensures the following:
For any positive integer $i\leq n$ and $j<n$, $$\|S_{\nyil e(n,n)+3(i-1)n+3j}\|= \| S_{\nyil e(i,j)}\| \cap \| S_{\nyil x(i)}\|$$ and
$$\|S_{\nyil e(n,n)+3in}\|= (\| S_{\nyil e(i,n)}\| \cap \| S_{\nyil x(i)}\|) \cup \overline{\| S_{\nyil x(i)}\|} .$$
With the notation $c(i,j) = e(n,n) +  3(i-1)n+3j$, the last two properties lead to the following statement.

\beginLemma\label{lem:4} If $r \in \|S_{\nyil x(i)}\|$ then ($r \in \|S_{\nyil c(i,j)}\|$ $\Leftrightarrow$ the $j$th bit of $(a^{2^i} \rmod p)$ is $1$), otherwise
$r \in \|S_{\nyil c(i,j)}\|$ $\Leftrightarrow$ $j=n$.
\endLemma

\begin{figure}
\centering
\includegraphics[width=0.9\textwidth]{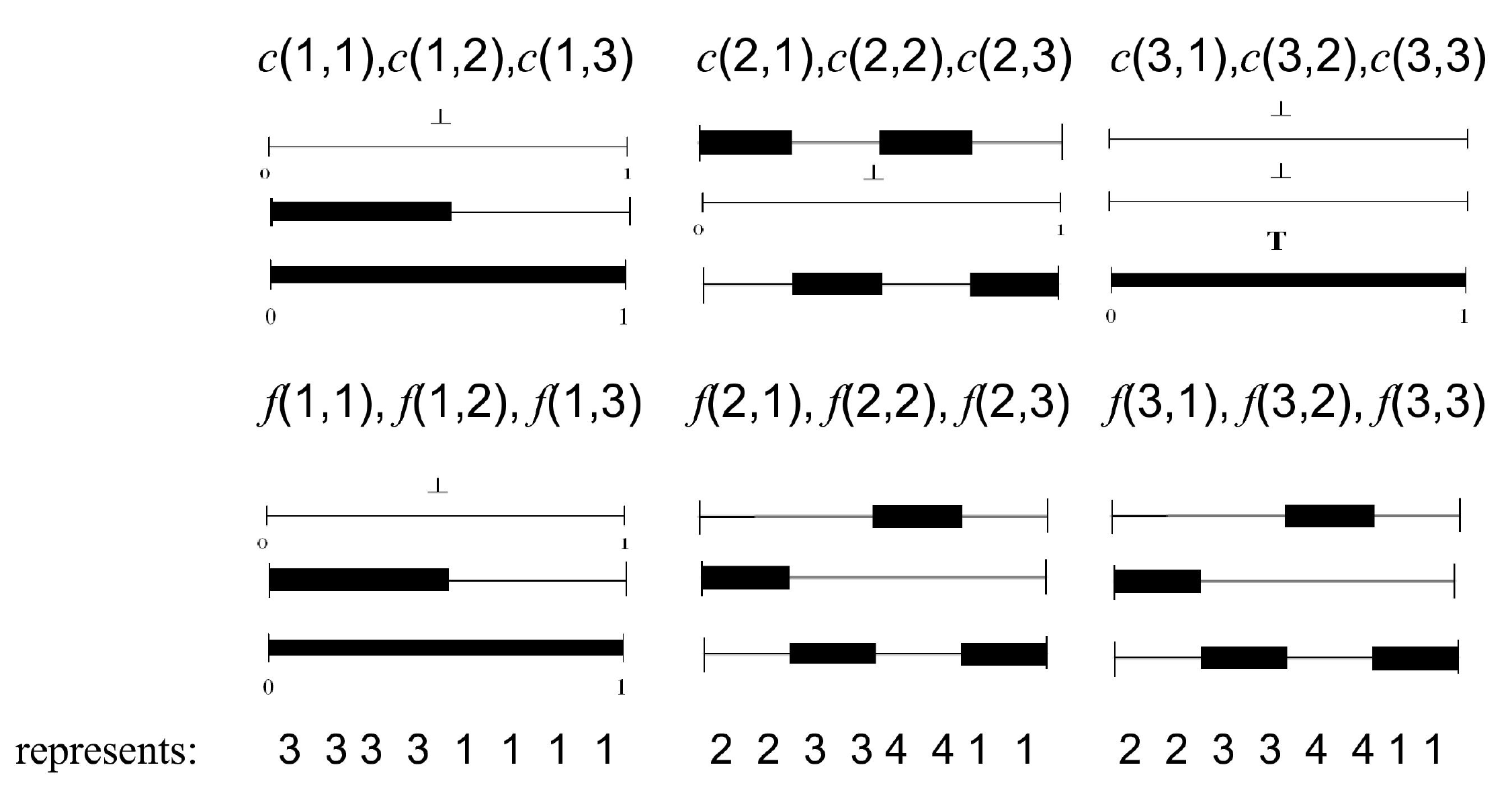}
\caption{The values $c(i,j)$ and $f(i,j)$ with $a=3$, $b=2$ and $p=5$.}\label{fig:cf}
\end{figure}

By reusing the circuit for multiplication, $\mathcal{B}$ continues the computation in such a way that the following requirement fulfills,
with the notation $f(i,j) = e(n,n) + (i-1)n + im +j$ \ \ ($1 \leq i,j\leq n$),
$$r \in \|S_{\nyil f(i,j)}\| \Leftrightarrow\mbox{ the $j$th bit of }\left(\prod\limits_{k=1}^{i}[x_k(r)(a^{2^k} \rmod p)] \rmod p\right).$$
The interval-values $c(i,j)$ and $f(i,j)$ of our example are shown in Figure \ref{fig:cf}. The next lemma is a direct corollary of Lemma \ref{lem:4}.

\beginLemma\label{lem:5}
For any possible inputs $a,p$ and for any $r \in [0,1)$, $\# (r \in \|S_{\nyil f(n,1)}\|,\ldots,r \in \|S_{\nyil f(n,n)}\| )$ is $a^{\#X(r)} \rmod p$.
\endLemma

From this point $\mathcal{B}$ continues with the equality test and output generation.
Equality test means pointwise checking for $\forall k \in \{1,\ldots,n\}: \  r \in \| S_{\nyil b(k)}\| \Leftrightarrow r \in \|S_{\nyil f(n,k)} \|$.
$f := f(n,n)$. The following computation will do it. For any $k \in \{1,\ldots,n\}$, let
\begin{align*}
S_{\nyil f+5k-4} &= (\AND,b(k),f(n,k)),\\
S_{\nyil f+5k-3} &= (\NOT,b(k)),\\
S_{\nyil f+5k-2} &= (\NOT,f(n,k)),\\
S_{\nyil f+5k-1} &= (\AND,f+5k-3,f+5k-2),\\
S_{\nyil f+5k} &= (\OR,f+5k-4,f+5k-1)\mbox{, and}\\
S_{\nyil f+5n+1} &= (\AND,f+5,f+5);
\intertext{for any  $k \in \{2,\ldots,n\}$, let}
S_{\nyil f+5n+k} &= (\AND,f+5k,f+5(k-1)).
\end{align*}
Now, let $e$ denote $f+6n$.

\beginLemma\label{correctness}
For any possible inputs $a,p,b$ and for any $r \in [0,1)$, $r \in \|S_{\nyil e}\| \Leftrightarrow  b=a^{\#X(r)} \rmod p$.\endLemma

\beginProof We observe that
\begin{align*}\forall k \in \{1,\ldots,n\}  \forall r \in [0,1): &r \in \|S_{\nyil f+5k}\| \Leftrightarrow ( r \in \|S_{\nyil b(k)}\| \Leftrightarrow r \in \|S_{\nyil f(n,k)}\|),\\
\forall k \in \{2,\ldots,n\}  \forall r \in [0,1):
&r \in \|S_{\nyil f+5n+k}\| \Leftrightarrow \forall j \in \{1,\ldots,k\}: ( r \in \|S_{\nyil b(j)}\| \Leftrightarrow r \in \|S_{\nyil f(n,j)}\|).
\end{align*}
For $k=n$, the last statement means that
$\forall r \in [0,1)$: $r \in \|S_{\nyil e}\| \Leftrightarrow \forall j \in \{1,\ldots,n\}: ( r \in \|S_{\nyil b(j)}\| \Leftrightarrow r \in \|S_{\nyil f(n,j)}\|)$, that is,
$b=a^{\#X(r)} \rmod p$ from  Lemma \ref{lem:5} and the fact, that $ b = \#(r \in  \|S_{\nyil b(1)}\|,\ldots,r \in  \|S_{\nyil b(n)}\|)$, independently from $r$.
\endProof

Now the proof of the main theorem is continued with separation of an $\frac{1}{2^n}$-size subinterval of $\|S_{\nyil e}\|$ that describes a %
solution. More
definitely, we find a subinterval $S$ of   $\|S_{\nyil e}\|$ that $\forall r,t \in S: \#X(r)=\#X(t)$ holds. It is an important step because $\|S_{\nyil e}\|$ may describe more (at most two) solutions.

First a 7-step process is used to separate the first component of $\|S_{\nyil e}\|$.
\begin{align*}
 e+1 :~ &(\NOT,e),\\
 e+2:~ &(\LSHIFT,e,e+1),\\
e+3:~&(\LSHIFT,e+2,e+2),\\
e+4:~ &(\RSHIFT,e+3,e+2), \\
e+5:~& (\RSHIFT,e+4,e+1), \\
e+6:~  &(\NOT,e+5), \\
e+7:~ &(\AND e, e+6).
\end{align*}
This computation guarantees that $\|S_{\nyil e+7}\|$ is the first component of $\|S_{\nyil e}\|$.
Based on the fact that the solution $x=0$ implies another solution $x<p$ that
is positive integer, we use right-shift to obtain an empty subinterval $\left[ 0, \frac{1}{2^n} \right)$.
\begin{align*}
e+8:~&(\RSHIFT,e+7,x(n)), \\
e+9:~&(\LSHIFT,e+8,x(n)), \\
e+10:~&(\RSHIFT,e+9,x(n)).
 \end{align*}
Then
\begin{align*}
 e+11:~&(\NOT,e+10), \\
 e+12:~&(\LSHIFT,e+10,e+11), \\
 e+13:~ &(\AND,e+12,x(n)), \\
 e+14:~&(\NOT,e+13), \\
 e+15:~&(\LSHIFT,e+13,e+14), \\
 e+16:~&(\LSHIFT,e+15,e+15), \\
 e+17:~ &(\RSHIFT,e+16,e+15), \\
 e+18:~& (\RSHIFT,e+17,e+14), \\
 e+19:~  &(\NOT,e+18), \\
 e+20:~ &(\AND,e+13, e+19), \\
 e+21:~ &(\RSHIFT,e+20,e+11).
 \end{align*}
Finally, it is shifted back to the correct place
\begin{align*}
e+22:~& (\LSHIFT,e+21,x(n)).
\end{align*}

\beginLemma
For all $r,t \in S_{e+22}$, we have $\#X(r)=\#X(t)$.
\endLemma

\beginProof
There are two possible cases. The first component of  $\|S_{\nyil e}\|$ is just an $\frac{1}{2^n}$-sized subinterval or longer. In both cases
$\|S_{\nyil e+12}\|$ is a left-shifted version of $\|S_{\nyil e+7}\|$. Even if it is longer than $\frac{1}{2^n}$, the first component of
$\|S_{\nyil e+13}\|$ has exactly  length $\frac{1}{2^n}$. $\|S_{\nyil e+21}\|$ is computed by the same way from $\|S_{\nyil e+13}\|$ as
$\|S_{\nyil e+7}\|$ from $\|S_{\nyil e}\|$, so $\|S_{\nyil e+21}\|$ is the first component of $\|S_{\nyil e+13}\|$. In $\|S_{\nyil e+22}\|$, this component is shifted back to its original place. %
 So $\|S_{\nyil e+22}\|$ is the left $\frac{1}{2^n}$-length prefix of the first component of $\|S_{\nyil e}\|$.
\endProof

\begin{figure}
\centering
\includegraphics[width=0.9\textwidth]{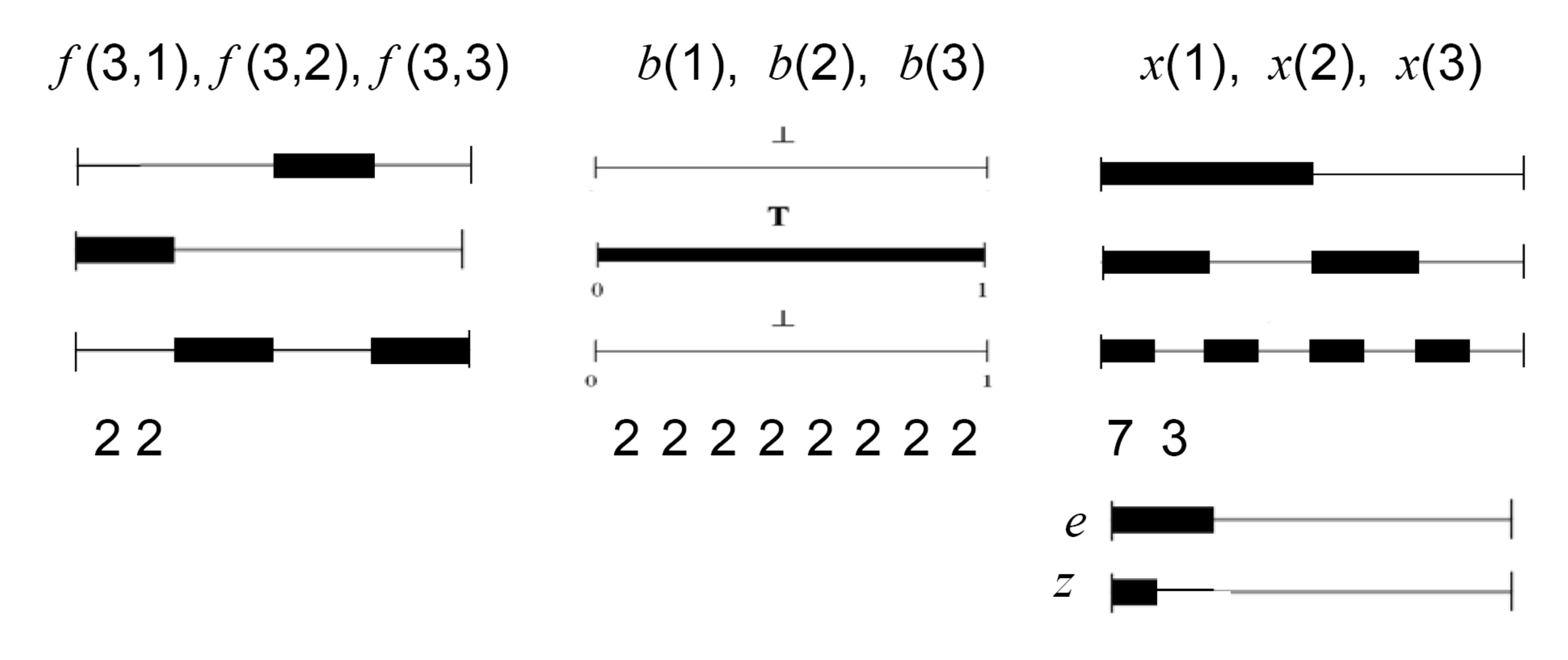}
\caption{The final part of the computation on input $a=3$, $b=2$ and $p=5$.}\label{fig:res}
\end{figure}
\medskip

Let $z$ denote $e+22$. Then $\mathcal{B}$ continues the computation sequence in the following way.
Let $S_{z+k}$ be $(\AND,z,x(k))$ for all $k \in \{1,\ldots,n\}$ and let $S_{z+n+k}$ be $(\OUTPUT,z+k)$. By the above results,
it is clear that for any possible input $a,b,p$, $\mathcal{B}$ will put out the bits of a solution $x$ of $a^x=b \rmod p$. That is, computing the discrete logarithm is finished. Analyzing the length of the computation validates that it is really in $O(n^3)$. 
In this way our main result, Theorem 1 is proved.

Let us continue our example. In Figure \ref{fig:res} some details of the final part of the computation is shown. The result 111 is written to the output representing the number 7.
One may easily check that $3^7 = 2187$ and $2187 \rmod 5 = 2$, the example computation is correct. With a small modification of the algorithm we may put the other value to
the output (if more than one solutions can be represented on $n$ bits).

\section{Concluding remarks}

In this paper, the efficiency of the interval-valued paradigm is presented by solving
the discrete logarithm problem by a cubic algorithm. Unfortunately our algorithm
uses high inner parallelism (high number of small interval segments in the unit interval) and therefore it does not help to solve the problem in traditional computers.

Based on some similarities of our new result  and the result presented in \cite{publi2011}), we could formulate
a conjecture: any function computation problem  that $\{(x,y) | f(x)=y \}$ is checkable on a Boolean circuit of polynomial size can be solved by a
interval-valued computation of polynomial size, where the computation has a special form:
the product operator is used only at the beginning, where `all possible inputs' are generated, and later product is not used. Moreover, it seems that the reverse direction of the conjecture also holds.

As we already mentioned a class of restricted polynomial size interval-valued
 computations characterizes PSPACE. It is an interesting challenge to analyse
 the power of non-restricted case and the relations of various classes of interval-valued computations to other paradigms, e.g., to vector
 machines \cite{vector}.

\section*{Acknowledgements}
The work is supported by the T\'AMOP 4.2.1/B-09/1/KONV-2010-0007 and
by the T\'AMOP 4.2.2/C-11/1/KONV-2012-0001
projects.
The projects are implemented through the New Hungary Development Plan, co-financed by the
European Social Fund and the European Regional Development Fund.

\bibliographystyle{eptcs}
\bibliography{nagy}
\end{document}